# VALUATION OF TRAVEL TIME SAVINGS IN THE PRESENCE OF SIMULTANEOUS ACTIVITIES

J. Pawlak
J.W. Polak
Centre for Transport Studies, Imperial College London

## 1. INTRODUCTION

Each day individuals undertake a number of various activities. In doing so, they decide on what to do when, where, for how long and in what order. As a result, apparently simple choices such as time of shopping or what to do on the train, as well as more difficult ones regarding hours of work or commuting mode, create a complex system of interrelated activity decisions. In fact, each individual seeks an arrangement of activities that is most convenient to their circumstances. These, on the other hand, may vary depending on numerous factors, e.g. socioeconomic characteristics. In practice, each person finds their most convenient pattern of activities through everyday experience, by iteratively adapting their behaviour and schedule. However, a picture of the current state of modelling reveals a shortcoming of the existing frameworks in treatment of activities which can be undertaken simultaneously, i.e. time sharing activities. While such activities form an important part of human life (working on a train, talking on the phone while shopping), they have received surprisingly little attention within the existing microeconomic frameworks. Moreover, time sharing possibility can have an impact on the estimates of valuation of time as a resource.

Hence, this study seeks to quantitatively explore factors which affect people's choices regarding time sharing activities. While theoretical part of this work generalises to any kind of time sharing activities and also discusses the conceptual issues present in that research field, the empirical part has been, due to the data availability, confined to investigating time sharing between travelling on a train and other activities. In particular, the following research objectives are aimed to be achieved:

- to develop a discrete choice model enabling investigation of how individuals choose to allocate their travel time between various combination of activities depending on journey- and individual-related factors;
- to calibrate and validate the developed model with the use of empirical data obtained from the National Rail Passenger Survey Autumn 2004 (SRA, 2004);
- to comparatively examine factors underlying the decisions of choosing certain activity or combination of activities while travelling on the train.

## 2. BACKGROUND AND RATIONALE

The concept of an *activity* has been a topic of interest in various disciplines for a long time. Whereas a number of modelling approaches looking at the problem from a perspective of utility maximisation have been developed, the concept of an *activity* also poses a number of difficulties for researchers.

The main problem arises from defining the concept in a clear and convenient, yet functional manner. While encyclopaedic definitions are available, they remain far from sufficient for the purpose of quantitative empirical research. Such definitions do not provide indication of the level of disaggregation of activities, e.g. should reading a book and reading a paper be considered separately or together as reading activity. Should sleeping or window-gazing be named activities or rather 'anti-activities' serving as transitory periods for 'shifting gears' (Mokhtarian and Salomon, 2001)? The problem of the aggregation level is also related to time sharing activities. In fact, if one defines each possible combination of activities undertaken simultaneously in a very short period as separate activities, the resulting disaggregation would mean that technically overlapping activities (Floro and Miles, 2003) are converted into singular ones. Moreover, even individuals undertaking seemingly the same physical activities can differ in their cognitive activities, e.g. gazing at the window and daydreaming or gazing at the window and working out a mathematical problem. From a modelling point of view, such a detailed disaggregation could cause problems in terms of collecting empirical data or numerous parameters' estimation. Thus, such an approach appears inconvenient in dealing with time sharing activities.

Another problem stems from the fact that activities are interlinked in various complex ways. As a result, one activity might be related to other activities in numerous, individual-specific ways: from stimulating (e.g. exercising and eating), being a precursor to (e.g. shopping and cooking) influencing the utility derived from a certain activity (e.g. walking in the rain having bought an umbrella), to limiting (e.g. eating and exercising afterwards) or preventing at all (e.g. buying coffee and entering library) (Jacobson et al., 2000; Mokhtarian, 2005; Melloni et al., 2007).

A third issue arises from the fact that an individual's behaviour can be impacted by the availability of information and communication technologies (ICT). ICT can affect the form in which certain activities are performed, e.g. traditional shopping versus mobile commerce (Hu, 2009). Also the extent to which activity can be undertaken under different conditions can vary as claimed by Andreev et al. (2007) who proposed a framework in which activity could be assessed in terms of its spatial, contextual and temporal flexibility. While the issue is important in the context of time sharing activities, such a perspective has not been identified by the authors.

So far the issue of time sharing activities has received a fairly limited treatment from the researchers. To the best of our knowledge, the most comprehensive empirical considerations have been done by Floro and Miles (2003) and Lyons et al. (2007). Floro and Miles used data from the Australian Time Use Survey to estimate Tobit models in order to explore the influence of various economic, demographic and social factors on what they called 'the incidence of overlapped work activity' (Floro and Miles, 2003). In doing so, they managed to establish that factors such as gender, household characteristics, education, cultural norms and level of income played a role as explanatory variables for the pattern of time sharing. However, their model was of a general exploratory character and did not relate to the issue of time valuation.

On the other hand, Lyons et al. (2007) investigated the issue from a more transport-oriented perspective by exploring time sharing time between activities undertaken by passengers travelling by train. As a result it has been revealed that 78% of respondents claimed they used their travel time productively by 78%. Jain and Lyons noted, however, that such behaviour is not a rail-specific phenomenon, yet the mode of transport can have significant impact on the pattern of time sharing (Jain and Lyons, 2008). Moreover, travel time can be also influential in terms of what activities are undertaken (Schwanen and Dijst, 2002).

Additionally, a survey conducted by British Telecom has revealed that significant proportion of business people take part in telephone conferences while on a train, ski slope or in a theme park (Telegraph Media, 2010). Nonetheless, so far no comprehensive modelling framework for the issue has been proposed while Fickling et al. (2008) called for one. This paper, aims therefore to answer that call by providing a modelling framework for exploring the productive use of travel time, which is effectively time sharing between travel and other activities.

From a modelling perspective, one of the most convenient tools for analysing activity behaviour proved to be time allocation models. Perception of time as a resource playing an important role in the process of utility maximisation of an individual began as early as in the nineteenth century with the works of Jevons (Jara-Diaz et al., 2009). In the 1960s the issue started to be explored again, following Gary Becker's proposition of a microeconomic model linking the utility derived from undertaking various activities with resource and time constraints (Becker, 1965). The most significant subsequent augmentations and reformulations of that model can be attributed to Johnson (1966), DeSerpa (1971), Evans (1972), Small (1982), and Jara-Diaz (2003). Apart from providing a convenient framework representing the process of time allocation as an optimisation problem, they frameworks enable estimation of the resource value of time, a concept 'which arises because total amount of time available for allocation to all activities is fixed by the total time constraint' (Hess et al., 2005). This parameter provides a basis for economic appraisal of transport investments. Thus the subsequent chapter shall provide a theoretical perspective on the possibility of introducing the idea of time sharing activities in time allocation models.

## 3. THEORETICAL APPROACH

Various aspects of time allocation models have received attention and critique from the academics, which resulted in different reformulations of the frameworks and also different expressions for resource value of time. However, the well-established form of time constraint (Jara-Diaz, 2007):

$$\sum_{i=1}^{n} T_i = \tau$$

where:

    $T_i$        time allocated to activity *i*
    *n*        set of all activities
    *τ*        total time available

has been particularly resistant to any critique and reformulation. While its form is convenient for establishing the clear and tractable optimisation problem, it implicitly assumes that certain amount of time (interval) can be allocated to only one activity, i.e. there is no time sharing. Modification of this well-established time constraint in a way enabling accommodation of time sharing activities, which could be practically interpreted as increased amount of the time available, can lead to a change in valuation of time as a resource (Kono et al., 2007). The issue is of a great importance, since the resulting resource value of time is used to estimate value of travel-time savings (VTTS), i.e. monetary value of the time saved from reduced travel time (Hess et al., 2005). However, acknowledging that travel time can be used productively, i.e. effectively, including the possibility of time sharing activities, may result in the VTTS being revised downwards (Mackie et al., 2003).

In order to make the time constraint capable of accommodating time sharing activities, the following reformulation is proposed (see Pawlak, 2010):

$$\sum_{i=1}^{n}\sum_{j=1}^{n}\cdots\sum_{y=1}^{n}\left(\frac{[z-k(T_{ij..y})+1]!}{z!}\right)\delta_{ij...y}T_{ij...y}=\tau$$

where:

    *i, j...y*    activities undertaken;
    *n*    total number of activities that an individual can undertake;
    *z*    maximum allowed number of activities sharing the same time interval;
    *k($T_{ij...y}$)*    number of different activities undertaken in the considered time interval;
    *$δ_{ij...y}$*    dummy variable which is equal to 1 if *i,j,...,y* combination of activities is feasible, and 0 otherwise;
    *$T_{ij...y}$*    time interval during which activities indicated by the lower indices (*i,j,...,y*) are undertaken *together* i.e. they share time interval;
    *τ*    total time available.

Moreover, in order to assure consistency it must be that:

$$n \geq z \geq k$$
$$n, k, z \; \epsilon \; N_+$$

which is a formal way of saying that number of activities undertaken in a given time interval (***k***) must be less or equal number of allowed time-sharing activities (***z***) which must not be greater than all possible activities (***n***). Finally, all these numbers must be positive natural numbers.

    Such a form of a time constraint can be understood as a summation of all the elements of the ***z***-dimensional symmetric tensor of ***n*** activities (and hence ***$n^z$*** elements) in such a way as to ensure that time intervals during which a particular combination of activities is undertaken is counted only once. This is because if more than one activity is undertaken in a given time interval, e.g. activities A and B, within the tensor this time interval will exist as elements ***$T_{AB}$*** and ***$T_{BA}$***. Without the parameter in the bracket, which depends on the number of possible combinations of certain number of activities, such time intervals would be double counted during summation. The assumption underlying this term is that all activities within the same time period are of equal importance, i.e. there is no primary activity. While such an assumption is conceptually debatable from, it does not alter the reasoning showing the influence of a change in time constraint on time valuation and hence shall be followed. Finally, the inclusion of dummy variable has been motivated by the fact that not all activities can share the same time period, e.g. it is impossible to drive a car and commute on the train at the same time. Thus, the dummy ensures preservation of a natural relationship between mutually exclusive activities during utility maximisation.

    It can also be observed that for ***z*** = 1, i.e. only one activity can be undertaken in a certain time interval, the time constraint reduces to a traditional form of time constraint. A more useful intuition can be provided with ***z*** = 2, i.e. maximum two activities share a given time period. In such a case, the time constraint can be understood as a summation of all elements of ***nxn*** matrix adjusted by the parameter preventing double-counting of the time intervals located outside the leading diagonal of the matrix. The idea of allocation of time to activities with the aid of matrix notation has been used by Evans (1972) to show the interdependence between activities.

However, as already noted, he did not consider the possibility of time sharing between activities. Following the notation above, such a matrix would be defined as:

$$\begin{pmatrix} \delta_{11}T_{11} & \cdots & \delta_{1n}T_{1n} \\ \vdots & \ddots & \vdots \\ \delta_{n1}T_{n1} & \cdots & \delta_{nn}T_{nn} \end{pmatrix}$$

whereas the resulting time constraint would be:

$$\sum_{i=1}^{n}\sum_{j=1}^{n}\left(\frac{[2-k(T_{ij})+1]!}{2!}\right)\delta_{ij}T_{ij} = \tau$$

It can be seen that time interval including activities 1 and *n* exists twice in the matrix (*T*$_{n1}$ and *T*$_{1n}$) and in order to ensure the consistency with the assumed total available time *τ*, the bracket term is essential. Such a form of time constraint can be introduced to the Small's time allocation model which includes three activities: work, leisure and time. For a greater clarity, a different notation consistent with that of Small (1982) is followed. In such a case the matrix describing time allocation is:

$$\begin{pmatrix} h & 0 & h^t(s) \\ 0 & l & l^t(s) \\ h^t(s) & l^t(s) & t(s) \end{pmatrix}$$

where:

| | |
|---|---|
| *h* | time spent exclusively on work |
| *l* | time spent exclusively on leisure |
| *t(s)* | time spent exclusively on travelling ('doing nothing') |
| *h$^t$(s)* | time shared between travel and work |
| *l$^t$(s)* | time shared between travel and leisure |

As can be noted, it has been assumed that leisure and work activities cannot share time, i.e. ***δ*$_{12}$ = *δ*$_{21}$** = 0, which is a reasonable assumption for a non-shirking individual. Also the three last terms are dependent on a schedule chosen by an individual. This is because the time schedule chosen determines the duration of travel and, as a result, time available for time sharing. If a time constraint based on such a matrix is introduced in Small's model (Small, 1982; Jara-Diaz, 2007), the resulting expression for the value of time has the following form (note that subscripts denote partial derivatives, e.g. ***A*$_B$** means partial derivative of ***A*** with respect to ***B***):

$$\frac{\mu}{\lambda} = \frac{1}{h_s^t + l_s^t + t_s}\left(wh_s^t + \frac{F_s(U_l - U_h - wU_x)}{F_h U_x} + \frac{U_s}{U_x} - c_s\right)$$

With the usual time constraint whereby **$h_s^t, l_s^t$** are equal to zero due to ***h$^t$*** and ***l$^t$*** being constrained to zero, the value of time is:

$$\frac{\mu}{\lambda} = \frac{1}{t_s}\left(\frac{F_s(U_l - U_h - wU_x)}{F_h U_x} + \frac{U_s}{U_x} - c_s\right)$$

where:

| | |
|---|---|
| *μ* | Lagrangean multiplier associated with time constraint |
| *λ* | Lagrangean multiplier associated with income constraint |
| *h* | time spent exclusively on work |
| *l* | time spent exclusively on leisure |
| *t* | time spent exclusively on travelling ('doing nothing') |
| $h^t$ | time shared between travel and work |
| $l^t$ | time shared between travel and leisure |
| *F* | function describing scheduling constraints |
| *U* | utility function of an individual |
| *x* | aggregate consumption |
| *w* | wage rate |
| *s* | time schedule |
| *c* | cost of travel |

While it is noticeable that formulations presented here differ from the forms presented by Small himself (Small, 1982; Jara-Diaz, 2007), they provide a good insight into potential differences in time valuation resulting from allowing for the existence of time sharing activities. In particular, two effects of the alteration in time constraint on the time valuation can be identified:

- Introduction of term $h^t_s + l^t_s$ in the denominator. These terms can be interpreted as the impact of scheduling on the amount of time spent on work and leisure while travelling. In other words, they measure the extent to which people react to changing travel circumstances. The impact of those terms is a reduction of the value of time as a resource since, with time sharing activities, it becomes a 'less scarce' resource.
- Introduction of term $wh^t_s$ inside the bracket. This can be interpreted as the extra money earned from working while travelling. This term, however, exists only if the extra amount of work undertaken while travelling translates directly into payment.

The above expressions show that time sharing activities can influence the valuation of time in two opposite ways. However, the direct translation of additional working time while travelling is dubious, especially given the differences between transport modes in terms of capabilities for travel activities undertaken e.g. working on a laptop is possible on a train, but difficult and illegal while driving a car. Thus the overall effect is likely to be such that the value of time as a resource (and thus also VTTS) is lower for estimation. This is consistent with the aforementioned suggestions of the need to revise time valuation by looking at travel time not only as an unproductive period as currently advised by the UK's Department for Transport's WebTAG (DfT, 2009), but also as an opportunity to undertake activities with positive utility (Mackie et al., 2003; Jain and Lyons, 2008).

Given the possible implications of inclusion of time sharing in time allocation frameworks on the resource value of time, it became crucial to explore what factors may influence the pattern of time sharing among individuals. Therefore, the empirical part of the research was aimed at investigating these factors.

## 4. EMPIRICAL STUDY

This study has been based on the data from the National Rail Passenger Survey Autumn 2004 (SRA, 2004). The data include responses from 25,596 passengers about their journey characteristics, socioeconomic status and, as an additional module in the Autumn 2004 wave, the type of activities undertaken while travelling on the train. However, to date the data have not been used to develop any model dealing with the travel activities and only Lyons et al. explored the data (Lyons et al. 2007), yet without developing any modelling framework.

**Table 1 Summary of the activity nesting structure**

| Nest | Activity |
|---|---|
| Passive time spending | Sleeping/snoozing<br>Window gazing/people watching<br>Being bored<br>Being anxious about the journey |
| Leisure (alone) | Reading for leisure<br>Playing games<br>Listening to music/radio |
| Interacting with other people | Talking to other passengers<br>Text messages/phone calls (personal)<br>Entertaining children |
| Work | Working/studying (reading/writing/typing/thinking)<br>Text messages/phone calls (work) |
| Other activities | Eating/drinking<br>Planning onward/return journey |

Source: SRA, 2004 (list of activities)

During the survey, each individual was able to select from as many as 14 different activities. In such a case the number of possible combinations of activities (any number and any mixture) in the choice set would be more than 16 thousand:

$$\sum_{k=1}^{14}\binom{14}{k} = 16\,383$$

Such a large number of alternatives would be virtually impossible to handle in a convenient manner. As a result, it has been decided to group the 14 elemental activities into 5 nests which results in 31 combinations (table 1).

In the context of modelling the choice of combinations of travel activities, the cross-nested logit (CNL) structure appeared the most appropriate. Firstly, it allows cross-nesting, i.e. alternatives can flexibly share unobservable characteristics, which is inevitable in modelling the choice from the set of combinations of activities. Secondly, as a member of the Generalised Extreme Value (GEV) models, it has a closed form expression for the probability. Thirdly, there exists readily available and robust software (BIOGEME 1.8) allowing estimation in a reasonable time.

In terms of the specification search, it had to be divided into two stages: an exploration of the available variables for explanatory power and construction of a complex model. Given there is a large number of potential explanatory variables, a widely appraised and frequently followed approach, so-called general-to-specific modelling (GETS) (Campos et al., 2005) was followed in the first stage. In the second stage, however, a trend alternative to the GETS was followed, namely simple-to-general approach (Hendry and Krolzig, 2001).

The procedure of validation involved the use of the $X^2$ test which enabled comparison of how many people in total are likely to undertake each particular combination of activities as calculated from the model and observed in the validation sample (20% of all observations). Hence, the hypothesis testing for goodness-of-fit could have been performed with null-hypothesis of no difference between the modelled and observed frequencies (Hogg and Ledolter, 1992). The result indicated to what extent the model provides a general description of the train travel activities.

## 5. RESULTS

In total 20,287 observations (estimation sample) were used to estimate 213 parameters: 30 combination-specific constants and 183 individual- and journey-related (table 3). The estimated models resulted in the $\rho^2$ value of 0.239 (0.236 if adjusted for the number of parameters). In terms of the error scaling parameters, a normalisation from the top means that the parameter $\mu$ remained fixed to 1, while the nest-specific parameters were estimated.

**Table 3 Summary statistics of the resulting CNL model specification**

| Statistic | Value |
|---|---|
| Number of estimated parameters | 213 |
| Number of observations | 20 287 |
| Null log-likelihood $\mathcal{L}(0)$ | -69 665.298 |
| Final log-likelihood $\mathcal{L}(\boldsymbol{\beta})$ | -53 023.425 |
| Log-likelihood ratio test | 33 283.747 |
| Rho-square $\rho^2$ | 0.239 |
| Adjusted rho-square $\bar{\rho}^2$ | 0.236 |
| Scaling parameter $\mu$ | 1 (fixed) |
| Scaling parameter for 'passive' nest $\mu_P$ | 1.81 |
| Scaling parameter for 'leisure' nest $\mu_L$ | 2.01 |
| Scaling parameter for 'interactive' nest $\mu_I$ | 1.42 |
| Scaling parameter for 'work' nest $\mu_W$ | 1.11* |
| Scaling parameter for 'other' nest $\mu_O$ | 1.07* |

*Insignificant at 95% level of significance

Table 4 summarises which variables were included in or rejected from the final model specification. Since the alternative specific constant for completely passive travel time spending was fixed to 0, positive parameters are associated with factors encouraging more active time spending. Items that were found to play a role were: newspaper, book, textbook, magazine, paperwork, games/puzzles, laptop, mobile phone, PDA, personal radio, game console and food. In case of the length of journey, the parameters were defined for individuals whose journeys lasted at least 1hour and associated with the *number* of activities undertaken. As far as gender is considered, the combination-specific parameters were defined for males. Other factors found significant and included in the complex model in an alternative specific form were: frequent use of the service (once a week), possibility of undertaking any paid work on the train, journey planning in advance, class of travel (first or second), trip's termination in central London, peak-time travel and availability of seating.

At the same time some variables were excluded from the final specification. Factors such as age, possession of luggage, bicycle, dog, wheelchair were found insignificant. Similarly, journey characteristics such as departure time, outward/return journey, length of use of the service in the past, typical trip over last month (seating and crowding conditions) or effect of weekend were also insignificant. In case of the companionship it was found that a variable representing any companionship performed better than a differentiation between adults and children. In terms of the working status or trip purpose, inclusion of the parameters representing these variables did not cause a significant improvement in the model's explanatory power.

**Table 4 Summary of the explored explanatory variables for the models**

| Variables included | |
|---|---|
| *Demographic* | *Companionship* |
| Sex | Travelling with any companion |
| *Equipment at hand* | *Journey characteristics* |
| Newspaper | Time spent on the train |
| Book | Frequency of using the service |
| Textbook | Possibility of working on the train (paid work) |
| Magazine | Journey planning in advance |
| Paperwork | First class traveller |
| Games/Puzzles | Trip terminating in central London |
| Laptop | Peak-time travel |
| Mobile phone | Availability of seating |
| PDA | |
| Personal radio | |
| Game console | |
| Food | |

| Variables rejected | |
|---|---|
| *Socioeconomic* | *Companionship* |
| Working status | Travelling with adult(s) |
| *Demographic* | Travelling with child(ren) |
| Age | *Journey characteristics* |
| *Equipment at hand* | Trip purpose |
| Luggage | Departure time |
| Bicycle | Outward/return journey |
| Dog | Length of use of the service in the past |
| Wheelchair | Typical trip over last month |
| | Weekday/Weekend |

The magnitude of impact of particular features of the individual and journey characteristics is presented in figure 1. While for a piece of equipment the value of the parameter itself was plotted, for the combination-specific parameters the median value and the inter-quartile range were included, so that the strength of the impact could have been compared. The strongest influence can thus be observed from the possession of certain equipment: paperwork (related to work activities), newspaper (related to work and leisure including activities), food (related to other activities), personal radio (related to leisure activities) mobile phone (related to work and interactive activities). Also being accompanied by someone, having pre-planned journey, class of travel and possibility of undertaking a paid work on the train were found influential.

The impact of having a book, textbook, laptop, PDA, games/puzzles or a game console was less significant. Similarly gender, termination of the journey at London's central terminus, peak travel, availability of seating or frequency of travel played smaller roles than the aforementioned factors. Interestingly, only the impact of trip's termination in London was consistent in discouraging travel activities for a completely passive time spending as indicated by the negative value of the associated parameter. The process of model validation was performed using a subsample of 5 049 individuals. Yet among these, only 844 observations included all responses necessary to calculate probabilities of choosing combinations of activities.

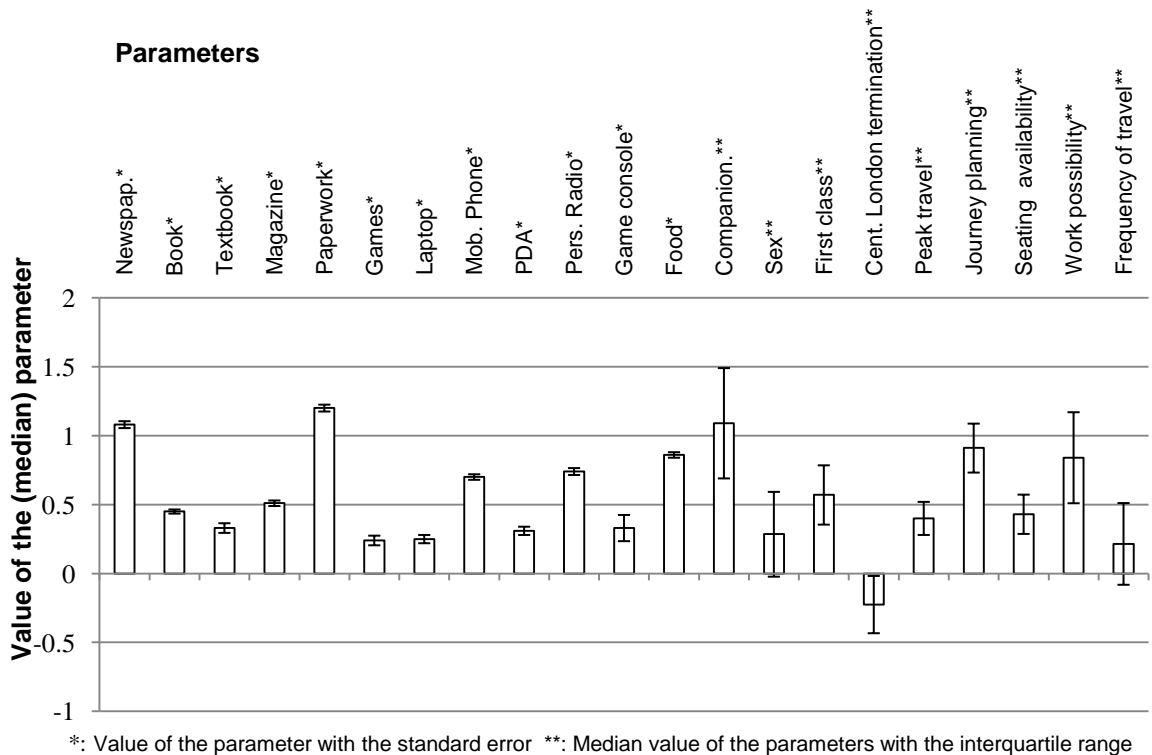

*: Value of the parameter with the standard error  **: Median value of the parameters with the interquartile range

**Figure 1 Comparative presentation of the model's parameter values**

The formal goodness-of-fit test $X^2$ revealed that the difference between the modelled and observed values is statistically significant, both for all combinations and for non-singular-activities combinations (6-31) at 95% and 99% levels of significance. However, as can be seen in figure 2, the match between the modelled and observed frequencies is still quite good, apart from passive time spending. Thus the model provides a reasonable description and exploratory tool of the data used for calibration, yet it is less robust in making predictions.

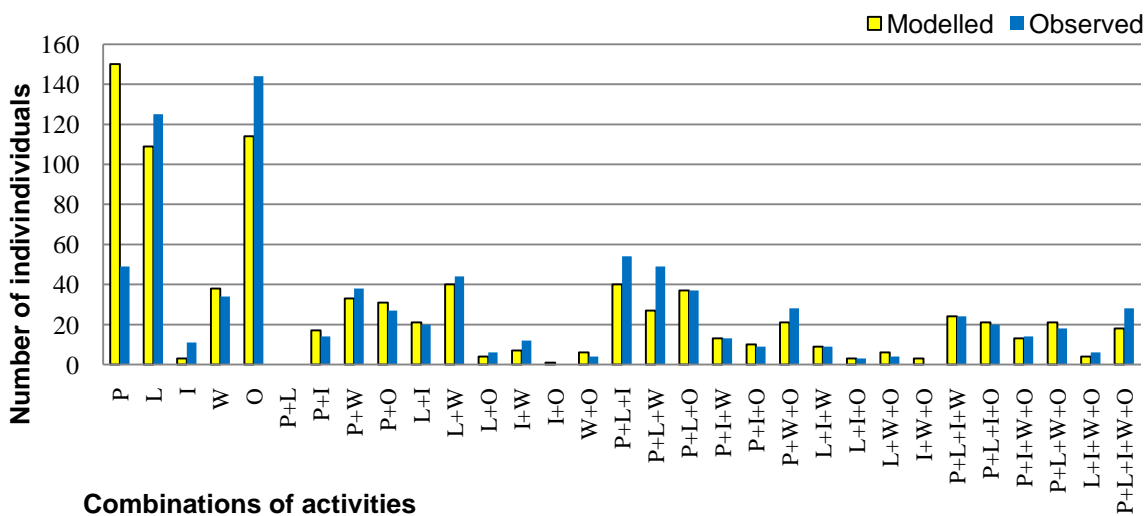

**Figure 2 Comparison of the modelled and observed frequencies in the validation sample**

# 6. DISCUSSION

Due to the way in which the research problem was defined, combinations of activities were inherently sharing certain features, having been generated as combinations obtained from a finite set of five nests (Passive, Leisure, Interactive, Work, Other). The strongest association of error components as indicated by the size of a nest-specific error scaling parameter was observed in 'Leisure', 'Passive' and 'Interactive' nests. This could be interpreted as greater similarity between the utilities of combinations including these activities as compared to 'Work' and 'Other' nests. In case of the 'Work' nest, the low value of the parameter can be interpreted as activities included in that nest (working, making work-related phone calls, studying) being not closely associated in terms of the utilities brought to individuals. In case of the 'Other' nest, statistical insignificance of the parameter may also result from the fact of the nest being quite artificial, including eating and planning journey. These essentially far different activities do not seem to bring similar utilities to individuals and hence the result is not surprising.

The model including only combination-specific constants reveals that in such a case the passive time spending is the most attractive alternative, while combinations involving work are the least. Unsurprisingly, possession of certain equipment was found significant in explaining choice of combinations where such a kit may be helpful, e.g. laptop in work or game console in leisure. Interestingly, the impact of ICT equipment is not significant as compared to newspapers, paperwork or food. However, at least some part of the explanation of such a situation may be attributed to the fact that the survey was undertaken in 2004 when wireless internet or 3G networks were not so widespread, possibly limiting the use of ICT. Items such as luggage, bicycle, wheelchair or dog were found as playing insignificant explanatory role, probably due to not being particularly good stimulants for travel activities on the train.

It has also been found that the fact of being accompanied by someone is a strong motivation for engaging in interactive activities with others. In case of the interactive activities, also gender plays an important role, which is also consistent with the results of Floro and Miles (2003). It was found that males are less interested in interactions but more interested in work-including combinations. On the other hand, age was found a statistically insignificant factor, with a possible reason being the fact that continuous variable was discretised leading to a numerically challenging estimation for age-specific parameter. In terms of the working status and trip purpose, the obtained parameters were significant, yet their inclusion in the final model did not improve its specification. As a result the parameter was dropped on the grounds that other variable(s) already represented its effect.

The remaining parameters that entered the model were related with the journey characteristics. It was found that pre-planning of travel activities was associated with higher propensity to engage in work, leisure or other activities. Interactive and passive time spending seemed more spontaneous. This could mean that planning in advance involve preparation e.g. taking certain equipment. Also possibility to undertake paid work on the travel was, as expected, a stimulus for work-related activities. Interestingly, first class travel and seating availability were not that influential. On the other hand, termination in central London was a factor generally discouraging non-work travel activities. This could be explained by the crowding on the commuter services and propensity for commuters to get involved with their job earlier. Such an interpretation would fit the suggestion of Mokhtarian and Salomon (2001) that commuting time is a transitory period for 'shifting gears'.

On the other hand, it is quite surprising that the length of using of certain service did not play role nor did a typical trip over the last month. We interpret this to mean that people in our sample do not adapt their activity pattern basing on the experience of travel conditions (standing, seating). This can either be a proof of a limited adaptive behaviour of train passengers or of a highly adaptive behaviour, e.g. being prepared for all possible cases. The most probable answer is a mixture of the two, given that some individuals reported planning their journey ahead.

The picture that emerges from the discussion above presents the complexity of the time sharing activities as represented by travel activities. Time sharing phenomenon is not a chaotic result of spontaneous decisions, but can be pre-planned and influenced by numerous factors. The process of developing a suitable modelling framework for such a complex proved challenging. Although the resulting specification is limited by the nature of the available data, it managed to encompass a number of factors within one framework and as such can be perceived successful.

## 7. CONCLUSIONS

This paper aimed at exploring the issue of time sharing activities in the context of travel activities by developing a suitable cross-nested logit model and calibrating it using empirical data on travel activities of train passengers. Theoretical consideration include in the paper has shown that inclusion of time sharing possibility could lead to reformulation of time allocation models and the resulting estimation of the resource value of time.

In the course of the empirical research, it was found that a number of individual- and journey-related factors can play a significant role in determining the choice of combinations of activities undertaken while travelling on the train.

While the model proved useful in exploring the available data and understanding the underlying relations, it was less successful in making predictions. However, it indicates that the issue of time sharing activities constitutes a complex entity and cannot be ignored, for instance in time valuation or investigations of time impact of ICT on activity patterns. In order to further investigate the issues, more research in the field appears warranted.